\documentclass[letter]{aa}  

\usepackage{graphicx}
\usepackage{deluxetable}
\usepackage{times,epsfig} 
\usepackage{mathrsfs}  
\usepackage{natbib}
\usepackage{amssymb}
\usepackage{longtable}
\bibpunct{(}{)}{;}{a}{}{,}
\usepackage[varg]{txfonts}
\usepackage[section]{placeins}
\usepackage[english]{babel}
\usepackage{url}
\usepackage{arydshln}
\usepackage{stfloats}
\usepackage{color}
\usepackage{amsmath}
\usepackage{soul}

\authorrunning{Ryder et al.}
\titlerunning{SN~1978K}
\begin{document}

   \title{SN 1978K: An evolved supernova outside 
          our Local Group detected at millimetre wavelengths}
   \author{S. D. Ryder\inst{1}
          \and
          R. Kotak\inst{2}
          \and
          I. A. Smith\inst{3}
          \and
          S. J. Tingay\inst{4,5}
          \and 
          E. C. Kool\inst{6,1}
          \and
          J. Polshaw\inst{2}
          }

   \institute{Australian Astronomical Observatory, P.O. Box 915,
              North Ryde, NSW 1670, Australia.\\ 
              \email{sdr@aao.gov.au}
         \and
             Astrophysics Research Centre, School of Mathematics
             and Physics, Queen's University Belfast, BT7 1NN, UK.
         \and 
             Department of Physics and Astronomy, Rice University,
             6100 South Main, MS-108, Houston, TX 77251-1892, USA.
         \and
             Istituto di Radioastronomia, INAF, via P. Gobetti,
             Bologna, 40129, Italy.
         \and
             International Centre for Radio Astronomy Research,
             Curtin University, Bentley, WA 6102, Australia.
         \and
             Department of Physics and Astronomy, Macquarie
             University, Sydney, NSW 2109, Australia.
             }

   \date{Received 21 Sep 2016; accepted 6 Oct 2016}

 
  \abstract{Supernova 1978K is one of the oldest-known examples
   of the class of Type~IIn supernovae that show evidence
   for strong interaction between the blast wave and a dense, pre-existing
   circumstellar medium. Here we report detections of SN~1978K at both 
   34~GHz and 94~GHz, 
   making it only the third extragalactic supernova (after SN~1987A and
   SN~1996cr) to be detected at late-times at these frequencies.
   We find SN~1978K to be $>$400 times more luminous
   than SN~1987A at millimetre wavelengths in spite of the roughly nine year
   difference in ages, highlighting the risk in adopting SN~1987A as a template
   for the evolution of core-collapse supernovae in general. 
   Additionally, from new VLBI
   observations at 8.4~GHz, we measure a deconvolved diameter for SN~1978K
   of $\sim$5~milli-arcsec, and a corresponding average expansion velocity of
   $<$1\,500~km~s$^{-1}$. These observations provide independent evidence of 
   an extremely dense circumstellar medium surrounding the progenitor star.
   }


   \maketitle


\section{Introduction}
Following the first detection of a supernova at radio wavelengths
close to fifty years ago \citep[SN~1970G;][]{Gottesman:72}, it was
soon widely realized that multi-frequency radio light curves can yield
information about the configuration, density, and prevailing
conditions of the circumstellar medium
\citep[e.g.,][]{Chevalier1982b,Weiler:86}. For cases where the radio
light curves are well sampled, this information in turn can be
translated into mass loss rates from the progenitor as a function of
time, thereby revealing the likely evolutionary stages the star must
have passed through prior to explosion
\citep[e.g.,][]{Sramek:84,Williams:02, Kotak:06}. Additionally,
carrying out very long baseline radio-interferometry (VLBI) on nearby
($\lesssim$ few Mpc) objects can spatially resolve the expanding
ejecta, uncovering underlying asymmetries, potentially due to a binary
companion \citep[e.g.,][]{Marcaide:94, Bartel:03}.

It is generally well established that the radio emission arises from a
combination of synchrotron self-absorption and free-free absorption as
the blast wave from the supernova encounters the surrounding medium
ionized by the initial high energy flash from the explosion, but
shaped by the progenitor system
\citep{Chevalier1982a,Chevalier1982b}. At early epochs, the optical
depth is large, suppressing the radio emission. As the shock moves
outwards, the optical depth drops, and its frequency dependence
results in the transition from optically thick to optically thin
occurring first at higher frequencies. Post peak, the emission
declines smoothly as a power-law, unless further dense material is
encountered, perhaps from earlier mass loss episodes.

Although several hundred supernovae (SNe) have now been detected at
radio wavelengths, the overwhelming majority of these are detections
at early times. This can be attributed to a combination of intrinsic
faintness and sensitivity of currently available instrumentation on
the one hand, and the lack of circumstellar material (CSM) at large
distances from the supernova that is dense enough to generate radio
emission on the other.



Supernova 1978K (\object{SN 1978K}) in the late-type barred spiral
galaxy NGC~1313 was only the second supernova to be detected and
recognized as a supernova from its radio emission. and the first from
its X-rays \citep{Ryder1993}. At peak, it would have been one of the
brightest radio supernovae yet seen. The Cepheid-based distance to
NGC~1313 is $4.61\pm0.21$\,Mpc \citep{Qing2015}, although estimates
based on the tip of the Red Giant Branch method
\citep[e.g.,][]{Jacobs2009} put it slightly closer.  
Regardless, SN~1978K is one of the closest supernovae undergoing
ejecta-circumstellar interaction (Type IIn).
It is one of a small number of SNe amenable to long term
multi-wavelength monitoring, and it remains bright at all wavelengths
almost four decades after explosion.


In this {\it Letter\/} we report new high frequency, and high spatial
resolution imaging of SN~1978K with the Australia Telescope Compact Array
(ATCA) and Long Baseline Array (LBA). 


\section{Observations and Results}

\subsection{ATCA observations}

SN~1978K was observed at 34~GHz and at 94~GHz with the ATCA in
September 2014.
The 34~GHz observations were
performed with the array in the 6B configuration, and the 94~GHz
observations with the array in the more compact H214
configuration. The observations were carried out in two bands, each
of 2~GHz bandwidth, centred either on 33 and 35~GHz, or on 93 and
95~GHz, using the Compact Array Broad-band Backend 
\citep[CABB:][]{Wilson2011} with each 2~GHz band consisting of
$2048\times1$~MHz channels. The parameters of each observing
session are listed in Table~\ref{t:obspar}.

   \begin{figure}
   \centering
   \includegraphics[width=\hsize]{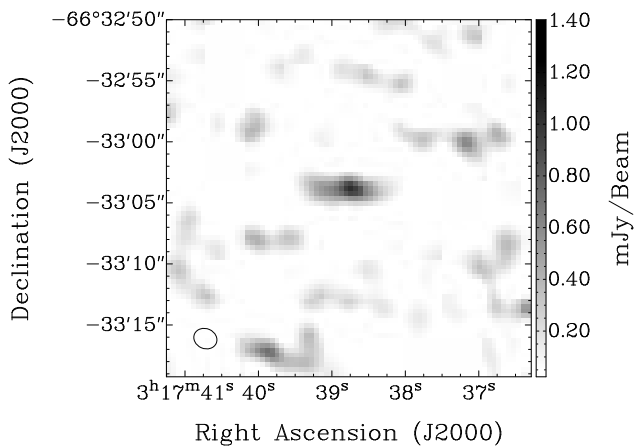}
      \caption{Continuum image of SN 1978K at 94~GHz from the
      ATCA on 2014 Sep 29, showing a 10$\sigma$ detection at the
      SN location (03$^{\rm h}$17$^{\rm m}$38.8$^{\rm s}$,
      $-66^{\circ}$33$^{\prime}$04$^{\prime\prime}$ J2000).
      The synthesised beam is indicated in the lower-left corner
      and the apparent source elongation is an artifact of the
      underlying noise pattern.} 
         \label{f:sn1978k_94}
   \end{figure}

\begin{table}
\centering
\caption{ATCA Observations of SN 1978K}\tabularnewline 
\vspace{0.2cm}
\label{t:obspar}\tabularnewline           
\begin{tabular}{l c c}
\hline\hline               

Parameter & 34 GHz & 94 GHz \\ 
\hline   
Date (UT)                 & 2014 Sep 14 & 2014 Sep 29 \\
Array configuration       &      6B     &    H214     \\
Max. baseline (m)         &     5969    &     240     \\
No. of antennas           &       6     &      5      \\
Total on-source time (hr) &      7.6    &     5.6     \\
Synthesised beam ($^{\prime\prime}$) &
                    $0.44 \times 0.24$  &  $1.9 \times 1.6$ \\
Image r.m.s. (mJy/beam)   &     0.03    &     0.13     \\
Source flux density (mJy) & $2.9\pm0.2$ & $1.2\pm0.3$ \\
\hline                                 
\end{tabular}
\begin{flushleft}
\end{flushleft}
\end{table}

The observations consisted of 12~minute scans on SN~1978K,
interleaved with 2~minute scans on the phase calibrator
PKS~B0302-623 having flux densities of $\sim$0.8\,Jy at 34~GHz and
$\sim$0.6\,Jy at 94~GHz. The sources PKS B2155-152 and
PKS B1921-293 served as the bandpass calibrators at 34~GHz
and at 94~GHz, respectively. Uranus was used as the primary
flux density calibrator at both frequencies, with hourly
pointing scans on the phase calibrator used to update the
pointing model. At 94~GHz "paddle" scans preceded each phase
calibrator observation to monitor the system temperatures.

The data were processed using the
{\sc miriad}\footnote{http://www.atnf.csiro.au/computing/software/miriad/}
package, and procedures outlined in Sec.~4.3 of the ATCA User
Guide\footnote{http://www.narrabri.atnf.csiro.au/observing/users\_guide/html/atug.html}.
The 34~GHz visibility data were corrected for atmospheric opacity,
while the 94~GHz visibility data had corrections applied to the antenna 
positions and system temperatures.
After editing and calibration of the data, images at each frequency
were made using a Briggs robust weighting factor of 0.5 as a best
compromise between resolution and sensitivity, then cleaned down to
three times the r.m.s. noise level. A clear detection of SN 1978K was
obtained in each frequency, and fitting of a Gaussian point source
yielded the flux densities shown in Table~\ref{t:obspar} where the
uncertainty in each case is a combination of the fitting and absolute
flux calibration errors. The resulting image of SN~1978K at 94~GHz is
shown in Fig.~\ref{f:sn1978k_94}. Core-collapse supernovae have
been detected previously at these frequencies
\citep[e.g., SN~2011dh:][]{Horesh2013}, but 
usually only within a few days or weeks after explosion.

\begin{figure}
   \centering
   \includegraphics[angle=-90,width=7cm]{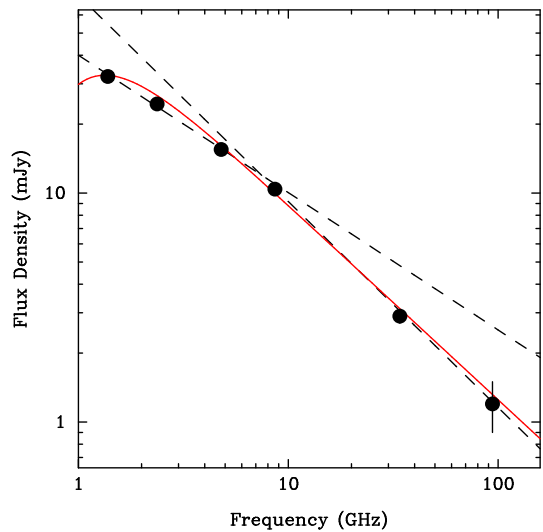}
      \caption{Radio spectral energy distribution for SN 1978K in
      September 2014. The dashed lines mark a power law slope of
      $-0.6$ passing through the $\nu < 10$~GHz data, and a
      steeper power law with slope of $-0.9$ passing through the
      $\nu > 10$~GHz data. The red solid line shows an SED
      fit to the full dataset using the model of \citet{Weiler2002},
      which includes low frequency absorption due to an
      H\,{\sc ii}~region along the line of sight.}
         \label{f:sed}
   \end{figure}

In Fig.~\ref{f:sed} we present the 1--100~GHz spectral energy
distribution of SN~1978K. The 1--10~GHz flux densities come from
ongoing monitoring with the ATCA, the most recent epoch for which is
2013~June~7~UT. The expected flux densities at 1.4, 2.4, 4.8, and
8.6~GHz for 2014~September have been derived using the model light
curve fits in \citet{Smith2007}. While a power-law index of
$\alpha=-0.6$ (where flux density $S_{\nu} \propto \nu^{\alpha}$)
provides a good fit to the data below 10~GHz, the power law steepens
to $\alpha\sim-0.9$ between the two higher frequency data points. We
note that even with the spectral steepening at high frequencies,
extrapolation of this power law to the regime of the Atacama Large
Millimetre/submillimetre Array (ALMA) Bands 4 (144~GHz) through 7
(325~GHz) should make the continuum emission from SN~1978K easily
detectable with ALMA.  
The Type~IIn SN~1996cr \citep{Meunier2013} in the Circinus Galaxy lies 
at a distance comparable to SN 1978K; it was found using the ATCA to have 
flux densities of $23.9 \pm 6.0$~mJy at 88.6~GHz at 3616 days after the 
explosion, and $40.3 \pm 6.0$~mJy at 34.5~GHz at 4842 days.
The flux densities measured recently for \object{SN~1987A} at
44 GHz \citep{Zanardo2013} and at 94 GHz \citep{Lakicevic2012} would,
if it were placed at the distance of NGC~1313 instead of 51.4~kpc
\citep{Panagia99}, correspond to $\sim$5~$\mu$Jy and $\sim$3~$\mu$Jy
respectively, {\em more than 2~orders of magnitude fainter than
  observed for SN~1978K}.


\subsection{VLBI observations}

SN 1978K was observed with the LBA on 2007 Nov 13 and
2015 March 29 UT. These gave similar results and we
discuss only the latter here because of the longer time
since explosion. For this observation the array was
comprised of 6 antennas: ATCA $5\times22$~m phased array;
Parkes 64~m; Mopra 22~m; Ceduna 30~m; Hobart 26~m; and a
single 12~m antenna of the Australian Square Kilometre Array
Pathfinder (ASKAP) in Western Australia, offering baselines
in excess of 3\,000~km. Although the observation block spanned
12~hr, the source was only above the elevation limit of some
antennas for $\sim$10~hr. A bandwidth of 64~MHz centred on
8.4~GHz was used in dual polarisation mode, and data recorded
to disk. The sources PKS B1921-293 and PKS B0537-441 served as
fringe-finders, while 3~minute scans of SN~1978K were alternated
with equivalent scans of the nearby source PKS B0355-669
($\sim$0.5~Jy at 8.4~GHz) to allow phase referencing.

\begin{figure}
   \centering
   \includegraphics[angle=-90,width=8cm]{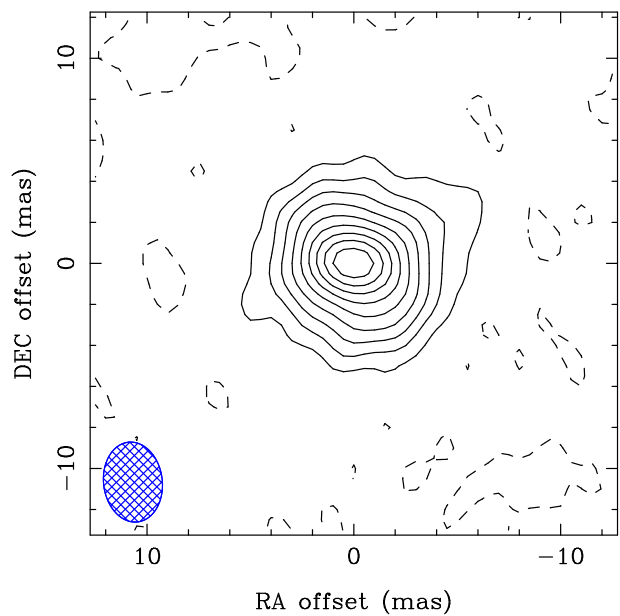}
      \caption{VLBI image of SN 1978K at a frequency of 8.4 GHz
      on 2015 March 29. The map is centered at 03:17:38.62,
      -66:33:03.4 (J2000). The beam shown at lower-left by the
      hashed ellipse is 3.9~mas $\times$ 2.8~mas. Contours are
      shown at $-2\%$, 10\%, 20\%, 30\%,$\ldots$,90\% of the peak
      brightness of 1.95~mJy/beam.}
         \label{f:vlbi}
   \end{figure}

All the data were correlated using the DiFX software
correlator \citep{Deller2011}, and the correlated data
were exported to AIPS\footnote{http://www.aips.nrao.edu/index.shtml}
for editing, applying nominal sensitivities for each antenna,
and iterative fringe-fitting.
Imaging using standard self-calibration, {\tt robust=0} weighting,
and clean techniques was carried out with the {\sc imagr} task.
The resultant image (Fig.~\ref{f:vlbi}) has a beam size of
$\sim$3~milli-arcsec and a dynamic range of $\sim$100:1. Deconvolution
of the VLBI source in the image plane using the {\sc jmfit} task
yields a deconvolved Gaussian size of 4.9~mas $\times$ 4.4~mas
($\pm0.5$~mas in each axis) at a position angle of
$(20\pm40)^{\circ}$ compared to the beam position angle of $85^{\circ}$. 
The integrated flux density of $5.2 \pm 0.3$~mJy is just half the
ATCA-measured value at 9~GHz in Fig.~\ref{f:sed}, suggesting that
some of the more-extended emission has been resolved out on account
of the longer baselines and sparse $uv$-coverage. Fitting of the visibilities
with a model in the $uv$-plane yielded similar results.


\section{Discussion}

\subsection{Radio SED}

While the 1--100~GHz radio spectrum of SN~1987A is consistent with a single
power law having $\alpha = -0.74$ \citep{Zanardo2013}, Fig.~\ref{f:sed}
shows this is clearly not the case for SN~1978K. A break such as this is
not unusual in core collapse supernovae at early phases, where synchrotron
self-absorption and/or free-free absorption results in a flattening of the
spectral index at lower frequencies that are still optically-thick.
However, after 36 years the expectation would be that SN~1978K should be 
optically thin at all these frequencies.

\citet{Montes1997} noted a similar curvature in the 0.8--9~GHz spectrum
of SN~1978K at $\sim$5\,000 days, which they attributed to an additional
time-independent free-free absorption component along the line-of-sight,
namely a foreground H\,{\sc ii}~region. Using their eqtn.~4 for the SED,
but with model parameters from our fit to the longer time baseline data
in \citet{Smith2007} we are able to reproduce the observed low frequency
turnover as shown in Fig.~\ref{f:sed} using a value for the thermal
optical depth parameter $K_{4}$ that is $2.5\times$ larger than that
determined by \citet{Montes1997}. This is nevertheless still consistent
with the presence of an intervening H\,{\sc ii}~region, or a circumstellar
ejecta nebula as argued by \citet{Chu1999} on the basis of optical
spectra. Observations at even lower frequencies would help to further
constrain this fit, but the GaLactic and
Extragalactic All-Sky MWA Survey \citep[GLEAM;][]{Wayth2015} at 70--230~MHz
currently lacks the necessary resolution to cleanly separate SN~1978K from its
host galaxy emission.

   \begin{figure}
   \centering
   \includegraphics[angle=-90,width=8cm]{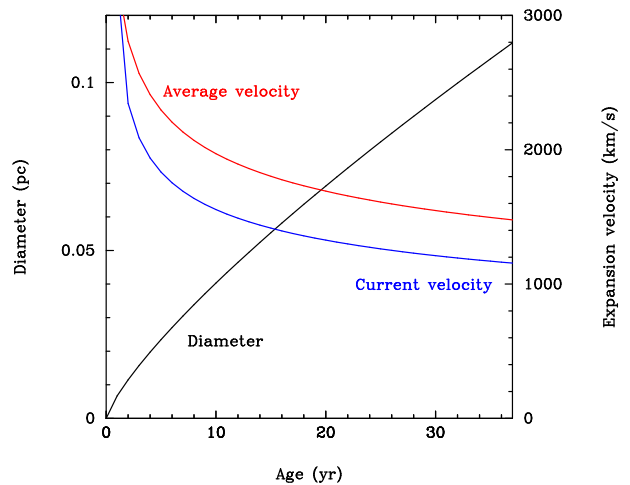}
      \caption{Modeled evolution of the remnant diameter of SN~1978K
      (black line), together with the current (blue line) and past average
      (red line) expansion velocities, for $R\sim t^{0.78}$.}
         \label{f:vd}
   \end{figure}

\subsection{Expansion velocity}

Were SN~1978K undergoing free expansion at a mean velocity of
10\,000~km~s$^{-1}$, it should span nearly 40~mas and be
easily resolvable by now. Our previous LBA observations at 1.4~GHz and
2.3~GHz of SN~1978K in 2003 \citep{Smith2007} put an upper limit on the
expanding remnant diameter of $\sim$10~mas, or 0.22~pc at the distance of
NGC~1313.
It appears that more than a decade later SN~1978K is still 
barely resolved even at 8.4~GHz,
with no apparent shell or ring-like structure yet visible as seen
in SN~1987A \citep{Zanardo2013} or SN~1993J \citep{Marcaide1993J}. We
can, however, halve the upper limit on the diameter to $<5$~mas (0.11~pc),
and consequently the average expansion velocity for SN~1978K
since explosion is $<$1\,500~km~s$^{-1}$. While much lower than
the 8\,000--10\,000~km~s$^{-1}$ average expansion velocity inferred from
VLBI measurements of objects like SN~1979C \citep{Marcaide1979c} 
and SN~1986J \citep{Bietenholz2010} some 25 years after their
explosions, we note the recently measured range of current expansion
velocities for SN~1987A of between 1\,600 and 6\,300~km~s$^{-1}$
depending on azimuth \citep{Zanardo2013}. This confirms that
the radio remnant of SN~1987A has also undergone significant deceleration
since the initial explosion at $\sim$35\,000~km~s$^{-1}$ \citep{Gaensler1997}.

We can make use of this new VLBI constraint on the current diameter of the
remnant of SN~1978K, together with information from the radio light curve
fits, to infer how the ejecta has decelerated over time.
The blast wave radius is assumed to grow with time as $R\sim t^{m}$, where
$m=1$ for no deceleration. In the \citet{Weiler2002} parameterisation of the
\citet{Chevalier1982a} model $m = -(\alpha - \beta - 3) / 3$, where $\alpha$ is
the radio spectral index, and $\beta$ is the rate of decline in flux density during
the post-peak, optically-thin phase. The SED fit shown in Fig~\ref{f:sed} has
$\alpha = -0.85$ and $\beta = -1.51$, so $m=-0.78$ which is at the extreme end
for decelerations measured this way \citep{Weiler2002}.

Fig.~\ref{f:vd} shows a model of the evolution in diameter, current expansion
velocity, and past average expansion velocity of SN~1978K for $R\sim t^{0.78}$
such that the current diameter is no more than 0.1~pc. Recent optical spectroscopy
of SN~1978K by \citet{Kuncarayakti2016} shows remarkable similarities with SN~1987A,
and line-widths indicating a current expansion velocity $\sim$500--600~km~s$^{-1}$,
within a factor of two of that predicted by this model. 
Whereas the radio emission at such late times is thought to trace the forward shock,
the optical emission lines are associated with the slower-moving reverse shock. 
\citet{Kuncarayakti2016} inferred an extreme mass-loss rate (0.01~$M_{\odot}$~yr$^{-1}$) 
based on the H$\alpha$ luminosity, leading to a massive circumstellar medium ($\sim$1\,$M_{\odot}$). Although there 
are several caveats that need to be considered when
converting H$\alpha$ luminosities to mass loss rates, the values above could 
simultaneously account for the significant radio luminosity as well as the dramatic 
deceleration in the expansion. 
%
Nevertheless we note that while Type~IIn supernovae are among the most luminous at
radio wavelengths, SN~1978K is in fact not unusually luminous for its type 
\citep[e.g.][]{Romero2014}. 


\section{Conclusions}

New radio observations of the Type IIn SN~1978K at high frequencies and at high
spatial resolution have led to some intriguing results:
\begin{enumerate}
\item SN~1978K turns out to be $>$400$\times$ brighter at frequencies
above 30 GHz than SN~1987A, 
the only other extragalactic supernova to have been 
detected at these frequencies more than 2~decades after explosion.
\item A single power-law does not fit the 1--100~GHz SED of SN~1978K. This may
be because the intrinsic emission has a broken power law, or due to the presence
of a nebular absorber along the line of sight as previously suspected from
low-frequency observations. Future metre-wave and sub-millimetre observations
should be able to discriminate between these two scenarios.
\item The remnant of SN~1978K is an order of magnitude smaller than expected
for the case of free expansion, and instead appears to be decelerating as
$R\sim t^{0.78}$. The past average expansion rate is below 1\,500~km~s$^{-1}$,
and the current expansion rate modeled to be 20\% lower still.
\end{enumerate}
These findings lend support to the arguments from contemporary X-ray and optical
analyses that the progenitor star of SN~1978K underwent extreme mass-loss in its
final centuries before explosion. This material is responsible for drastically
decelerating the expansion of the remnant in much the same way as observed in
SN~1987A, which shows other similarities with the older SN~1978K despite not
being classed as a Type~IIn event. 

\begin{acknowledgements}
SR and RK acknowledge support from the Royal Society International
Exchange scheme (IE140343). SR was a Visiting Research Fellow at
Queen's University Belfast where much of this paper was written.
RK also acknowledges support from STFC via ST/L000709/1.
We are grateful to Cormac Reynolds for invaluable assistance with
reducing the LBA data. We thank F.~Bauer for bringing to our 
attention the millimetre observations of SN 1996cr.
The Australia Telescope Compact Array and
Long Baseline Array are part of the Australia Telescope National
Facility which is funded by the Commonwealth of Australia for
operation as a National Facility managed by CSIRO. The
data reported here were obtained under Programs C184 and V157 
(P.I. S. Ryder). 
\end{acknowledgements}


\bibliographystyle{aa} 
\bibliography{sn1978k} 

\begin{thebibliography}{30}
\expandafter\ifx\csname natexlab\endcsname\relax\def\natexlab#1{#1}\fi

\bibitem[{{Bartel} \& {Bietenholz}(2003)}]{Bartel:03}
{Bartel}, N. \& {Bietenholz}, M.~F. 2003, \apj, 591, 301

\bibitem[{{Bietenholz} {et~al.}(2010){Bietenholz}, {Bartel}, \&
  {Rupen}}]{Bietenholz2010}
{Bietenholz}, M.~F., {Bartel}, N., \& {Rupen}, M.~P. 2010, \apj, 712, 1057

\bibitem[{{Chevalier}(1982{\natexlab{a}})}]{Chevalier1982a}
{Chevalier}, R.~A. 1982{\natexlab{a}}, \apj, 258, 790

\bibitem[{{Chevalier}(1982{\natexlab{b}})}]{Chevalier1982b}
{Chevalier}, R.~A. 1982{\natexlab{b}}, \apj, 259, 302

\bibitem[{{Chu} {et~al.}(1999){Chu}, {Caulet}, {Montes}, {Panagia}, {Van Dyk},
  \& {Weiler}}]{Chu1999}
{Chu}, Y.-H., {Caulet}, A., {Montes}, M.~J., {et~al.} 1999, \apjl, 512, L51

\bibitem[{{Deller} {et~al.}(2011){Deller}, {Brisken}, {Phillips}, {Morgan},
  {Alef}, {Cappallo}, {Middelberg}, {Romney}, {Rottmann}, {Tingay}, \&
  {Wayth}}]{Deller2011}
{Deller}, A.~T., {Brisken}, W.~F., {Phillips}, C.~J., {et~al.} 2011, \pasp,
  123, 275

\bibitem[{{Gaensler} {et~al.}(1997){Gaensler}, {Manchester}, {Staveley-Smith},
  {Tzioumis}, {Reynolds}, \& {Kesteven}}]{Gaensler1997}
{Gaensler}, B.~M., {Manchester}, R.~N., {Staveley-Smith}, L., {et~al.} 1997,
  \apj, 479, 845

\bibitem[{{Gottesman} {et~al.}(1972){Gottesman}, {Broderick}, {Brown},
  {Balick}, \& {Palmer}}]{Gottesman:72}
{Gottesman}, S.~T., {Broderick}, J.~J., {Brown}, R.~L., {Balick}, B., \&
  {Palmer}, P. 1972, \apj, 174, 383

\bibitem[{{Horesh} {et~al.}(2013){Horesh}, {Stockdale}, {Fox}, {Frail},
  {Carpenter}, {Kulkarni}, {Ofek}, {Gal-Yam}, {Kasliwal}, {Arcavi}, {Quimby},
  {Cenko}, {Nugent}, {Bloom}, {Law}, {Poznanski}, {Gorbikov}, {Polishook},
  {Yaron}, {Ryder}, {Weiler}, {Bauer}, {Van Dyk}, {Immler}, {Panagia},
  {Pooley}, \& {Kassim}}]{Horesh2013}
{Horesh}, A., {Stockdale}, C., {Fox}, D.~B., {et~al.} 2013, \mnras, 436, 1258

\bibitem[{{Jacobs} {et~al.}(2009){Jacobs}, {Rizzi}, {Tully}, {Shaya},
  {Makarov}, \& {Makarova}}]{Jacobs2009}
{Jacobs}, B.~A., {Rizzi}, L., {Tully}, R.~B., {et~al.} 2009, \aj, 138, 332

\bibitem[{{Kotak} \& {Vink}(2006)}]{Kotak:06}
{Kotak}, R. \& {Vink}, J.~S. 2006, \aap, 460, L5

\bibitem[{{Kuncarayakti} {et~al.}(2016){Kuncarayakti}, {Maeda}, {Anderson},
  {Hamuy}, {Nomoto}, {Galbany}, \& {Doi}}]{Kuncarayakti2016}
{Kuncarayakti}, H., {Maeda}, K., {Anderson}, J.~P., {et~al.} 2016, \mnras, 458,
  2063

\bibitem[{{Laki{\'c}evi{\'c}} {et~al.}(2012){Laki{\'c}evi{\'c}}, {Zanardo},
  {van Loon}, {Staveley-Smith}, {Potter}, {Ng}, \& {Gaensler}}]{Lakicevic2012}
{Laki{\'c}evi{\'c}}, M., {Zanardo}, G., {van Loon}, J.~T., {et~al.} 2012, \aap,
  541, L2

\bibitem[{{Marcaide} {et~al.}(1994){Marcaide}, {Alberdi}, {Elosegui},
  {Guirado}, {Mantovani}, {Perez}, {Ratner}, {Rius}, {Rogers}, {Schmidt},
  {Shapiro}, \& {Whitney}}]{Marcaide:94}
{Marcaide}, J.~M., {Alberdi}, A., {Elosegui}, P., {et~al.} 1994, \apjl, 424,
  L25

\bibitem[{{Marcaide} {et~al.}(2009{\natexlab{a}}){Marcaide},
  {Mart{\'{\i}}-Vidal}, {Alberdi}, {P{\'e}rez-Torres}, {Ros}, {Diamond},
  {Guirado}, {Lara}, {Shapiro}, {Stockdale}, {Weiler}, {Mantovani}, {Preston},
  {Schilizzi}, {Sramek}, {Trigilio}, {van Dyk}, \& {Whitney}}]{Marcaide1993J}
{Marcaide}, J.~M., {Mart{\'{\i}}-Vidal}, I., {Alberdi}, A., {et~al.}
  2009{\natexlab{a}}, \aap, 505, 927

\bibitem[{{Marcaide} {et~al.}(2009{\natexlab{b}}){Marcaide},
  {Mart{\'{\i}}-Vidal}, {Perez-Torres}, {Alberdi}, {Guirado}, {Ros}, \&
  {Weiler}}]{Marcaide1979c}
{Marcaide}, J.~M., {Mart{\'{\i}}-Vidal}, I., {Perez-Torres}, M.~A., {et~al.}
  2009{\natexlab{b}}, \aap, 503, 869

\bibitem[{{Meunier} {et~al.}(2013){Meunier}, {Bauer}, {Dwarkadas},
  {Koribalski}, {Emonts}, {Hunstead}, {Campbell-Wilson}, {Stockdale}, \&
  {Tingay}}]{Meunier2013}
{Meunier}, C., {Bauer}, F.~E., {Dwarkadas}, V.~V., {et~al.} 2013, \mnras, 431,
  2453

\bibitem[{{Montes} {et~al.}(1997){Montes}, {Weiler}, \& {Panagia}}]{Montes1997}
{Montes}, M.~J., {Weiler}, K.~W., \& {Panagia}, N. 1997, \apj, 488, 792

\bibitem[{{Panagia}(1999)}]{Panagia99}
{Panagia}, N. 1999, in IAU Symposium, Vol. 190, New Views of the Magellanic
  Clouds, ed. Y.-H. {Chu}, N.~{Suntzeff}, J.~{Hesser}, \& D.~{Bohlender}, 549

\bibitem[{{Qing} {et~al.}(2015){Qing}, {Wang}, {Liu}, \& {Yoachim}}]{Qing2015}
{Qing}, G., {Wang}, W., {Liu}, J.-F., \& {Yoachim}, P. 2015, \apj, 799, 19

\bibitem[{{Romero-Ca{\~n}izales} {et~al.}(2014){Romero-Ca{\~n}izales},
  {Herrero-Illana}, {P{\'e}rez-Torres}, {Alberdi}, {Kankare}, {Bauer}, {Ryder},
  {Mattila}, {Conway}, {Beswick}, \& {Muxlow}}]{Romero2014}
{Romero-Ca{\~n}izales}, C., {Herrero-Illana}, R., {P{\'e}rez-Torres}, M.~A.,
  {et~al.} 2014, \mnras, 440, 1067

\bibitem[{{Ryder} {et~al.}(1993){Ryder}, {Staveley-Smith}, {Dopita}, {Petre},
  {Colbert}, {Malin}, \& {Schlegel}}]{Ryder1993}
{Ryder}, S., {Staveley-Smith}, L., {Dopita}, M., {et~al.} 1993, \apj, 416, 167

\bibitem[{{Smith} {et~al.}(2007){Smith}, {Ryder}, {B{\"o}ttcher}, {Tingay},
  {Stacy}, {Pakull}, \& {Liang}}]{Smith2007}
{Smith}, I.~A., {Ryder}, S.~D., {B{\"o}ttcher}, M., {et~al.} 2007, \apj, 669,
  1130

\bibitem[{{Sramek} {et~al.}(1984){Sramek}, {Panagia}, \& {Weiler}}]{Sramek:84}
{Sramek}, R.~A., {Panagia}, N., \& {Weiler}, K.~W. 1984, \apjl, 285, L59

\bibitem[{{Wayth} {et~al.}(2015){Wayth}, {Lenc}, {Bell}, {Callingham},
  {Dwarakanath}, {Franzen}, {For}, {Gaensler}, {Hancock}, {Hindson},
  {Hurley-Walker}, {Jackson}, {Johnston-Hollitt}, {Kapi{\'n}ska}, {McKinley},
  {Morgan}, {Offringa}, {Procopio}, {Staveley-Smith}, {Wu}, {Zheng}, {Trott},
  {Bernardi}, {Bowman}, {Briggs}, {Cappallo}, {Corey}, {Deshpande}, {Emrich},
  {Goeke}, {Greenhill}, {Hazelton}, {Kaplan}, {Kasper}, {Kratzenberg},
  {Lonsdale}, {Lynch}, {McWhirter}, {Mitchell}, {Morales}, {Morgan}, {Oberoi},
  {Ord}, {Prabu}, {Rogers}, {Roshi}, {Shankar}, {Srivani}, {Subrahmanyan},
  {Tingay}, {Waterson}, {Webster}, {Whitney}, {Williams}, \&
  {Williams}}]{Wayth2015}
{Wayth}, R.~B., {Lenc}, E., {Bell}, M.~E., {et~al.} 2015, \pasa, 32, e025

\bibitem[{{Weiler} {et~al.}(2002){Weiler}, {Panagia}, {Montes}, \&
  {Sramek}}]{Weiler2002}
{Weiler}, K.~W., {Panagia}, N., {Montes}, M.~J., \& {Sramek}, R.~A. 2002,
  \araa, 40, 387

\bibitem[{{Weiler} {et~al.}(1986){Weiler}, {Sramek}, \& {Panagia}}]{Weiler:86}
{Weiler}, K.~W., {Sramek}, R.~A., \& {Panagia}, N. 1986, Science, 231, 1251

\bibitem[{{Williams} {et~al.}(2002){Williams}, {Panagia}, {Van Dyk}, {Lacey},
  {Weiler}, \& {Sramek}}]{Williams:02}
{Williams}, C.~L., {Panagia}, N., {Van Dyk}, S.~D., {et~al.} 2002, \apj, 581,
  396

\bibitem[{{Wilson} {et~al.}(2011){Wilson}, {Ferris}, {Axtens}, {Brown},
  {Davis}, {Hampson}, {Leach}, {Roberts}, {Saunders}, {Koribalski}, {Caswell},
  {Lenc}, {Stevens}, {Voronkov}, {Wieringa}, {Brooks}, {Edwards}, {Ekers},
  {Emonts}, {Hindson}, {Johnston}, {Maddison}, {Mahony}, {Malu}, {Massardi},
  {Mao}, {McConnell}, {Norris}, {Schnitzeler}, {Subrahmanyan}, {Urquhart},
  {Thompson}, \& {Wark}}]{Wilson2011}
{Wilson}, W.~E., {Ferris}, R.~H., {Axtens}, P., {et~al.} 2011, \mnras, 416, 832

\bibitem[{{Zanardo} {et~al.}(2013){Zanardo}, {Staveley-Smith}, {Ng},
  {Gaensler}, {Potter}, {Manchester}, \& {Tzioumis}}]{Zanardo2013}
{Zanardo}, G., {Staveley-Smith}, L., {Ng}, C.-Y., {et~al.} 2013, \apj, 767, 98

\end{thebibliography}

\end{document}